\RequirePackage{ifpdf}
\ifpdf 
\documentclass[pdftex]{sigma}
\else
\documentclass{sigma}
\fi

\begin{document}

\allowdisplaybreaks

\renewcommand{\PaperNumber}{037}

\renewcommand{\thefootnote}{$\star$}

\FirstPageHeading

\ShortArticleName{An Explicit Formula for Symmetric Polynomials}

\ArticleName{An Explicit Formula for Symmetric Polynomials
\\Related to the Eigenfunctions\\ of Calogero--Sutherland
Models\footnote{This paper is a contribution to the Proceedings of
the Workshop on Geometric Aspects of Integ\-rable Systems
 (July 17--19, 2006, University of Coimbra, Portugal).
The full collection is available at
\href{http://www.emis.de/journals/SIGMA/Coimbra2006.html}{http://www.emis.de/journals/SIGMA/Coimbra2006.html}}}

\Author{Martin HALLN\"AS} \AuthorNameForHeading{M. Halln\"as}
\Address{Department of Theoretical Physics, Albanova University
Center, SE-106 91 Stockholm, Sweden}
\Email{\href{mailto:hallnas@kth.se}{hallnas@kth.se}}
\URLaddress{\url{http://theophys.kth.se/~martin/}}

\ArticleDates{Received November 01, 2006, in f\/inal form February
05, 2007; Published online March 01, 2007}

\Abstract{We review a recent construction of an explicit analytic
series representation for symmetric polynomials which up to a
groundstate factor are eigenfunctions of Calogero--Sutherland type
models. We also indicate a generalisation of this result to
polynomials which give the eigenfunctions of so-called `deformed'
Calogero--Sutherland type models.}


\Classification{81Q05; 35Q58; 33C52}

\section{Introduction}
The purpose of this paper is to give a pedagogical review of a
recent construction of an explicit analytic series representation
for symmetric polynomials which up to a groundstate factor are
eigenfunctions of Calogero--Sutherland type models. In special
cases, this construction has been previously studied in
\cite{HallLang1,Lang1,Lang2}, while a detailed account of the more
general results presented in this paper is in preparation
\cite{HallLang2}.

To establish a context for our construction we begin this
introduction by brief\/ly discussing quantum many-body models of
Calogero--Sutherland type in general and highlighting some of the
distinguishing features of those models which have eigenfunctions
given by polynomials. By reviewing Sutherland's original solution
method for the Calogero model \cite{Suth1} we proceed to recall
that these polynomials have a triangular structure and to discuss
its importance when explicitly constructing them. We then sketch
the main steps in our solution method and compare it to
Sutherland's. An outline for the remainder of the paper f\/inally
concludes the introduction.

\subsection[Quantum many-body models of Calogero-Sutherland type]{Quantum many-body models of Calogero--Sutherland type}

A quantum many-body model of Calogero--Sutherland type is for some
potential functions $V$ and $W$ def\/ined by the Schr\"odinger
operator
\begin{gather}
\label{genHam}
  H_N = -\sum\limits_{j=1}^N\frac{\partial^2}{\partial x_j^2} +
  \sum\limits_{j=1}^NV(x_j) + \sum\limits_{j<k}W(x_j,x_k),
\end{gather}
where $N$ refers to the number of particles present in the system,
and $x_j$ to their positions. As f\/irst observed by Calogero
\cite{Calogero2} and Sutherland \cite{Suth1} in two special cases
there exist certain choices of these potential functions for which
the resulting Schr\"odinger operator can be exactly solved. For
many of these choices this is due to the fact that its
eigenfunctions are given by natural many-variable generalisations
of polynomials solving a second order ordinary dif\/ferential
equation. This includes for example the original models of
Calogero and Sutherland, whose eigenfunctions respectively
correspond to the Hermite polynomials and the ordinary monomials
$p_n = x^n$. We also mention Olshanetsky and Perelomov's
\cite{OlsPer} root system generalisations of these models, where
the Legendre, Gegenbauer and Jacobi polynomials similarly appear;
see \cite{BF,vanDie}.

All of these models share a number of remarkable properties: their
square integrable eigenfunctions, labelled by partitions $\lambda
= (\lambda_1,\ldots,\lambda_N)$, i.e. integers $\lambda_i$ such
that $\lambda_1\geq\cdots\geq \lambda_N\geq 0$, are all of the
form
\begin{gather}
\label{genEigFuncs}
  \Psi_\lambda(x_1,\ldots,x_N) =
  \Psi_0(x_1,\ldots,x_N)P_\lambda(z(x_1),\ldots, z(x_N))
\end{gather}
with particular symmetric polynomials $P_\lambda$ and a ground
state $\Psi_0$ which always is of the product form
\begin{gather}
\label{genGroundstate}
  \Psi_0(x_1,\ldots,x_N) =
  \prod\limits_{j=1}^N\psi_0(x_j)\prod\limits_{j<k}(z(x_k) - z(x_j))^\kappa,
\end{gather}
where the function $z$ is f\/ixed by the choice of potential
function $V$, and $\psi_0$ is the ground state of the one-body
model obtained by setting $N=1$ in \eqref{genHam}. The
corresponding eigenvalues are in addition of a very simple form
and can be written down explicitly. In the remainder of this paper
we will refer to the polynomials $P_\lambda$ as reduced
eigenfunctions of the corresponding Schr\"odinger operator
\eqref{genHam} and our aim is to explain an explicit analytic
series representation for them, obtained in \cite{HallLang2}.

\subsection{Triangular structures and Sutherland's solution method}
If we conjugate the Schr\"odinger operator \eqref{genHam} by its
groundstate \eqref{genGroundstate}, and subtract the corresponding
eigenvalue $E_0$, we obtain the dif\/ferential operator
\begin{gather}
\label{reducedGenHam}
  \tilde H_N := \Psi_0^{-1}(H - E_0)\Psi_0 =
  -\sum\limits_{j=1}^N\frac{\partial^2}{\partial x_j^2} -
  2\sum\limits_{j=1}^N\Psi^{-1}_0\frac{\partial\Psi_0}{\partial
  x_j}\frac{\partial}{\partial x_j},
\end{gather}
which has the symmetric polynomials $P_\lambda$ as eigenfunctions.
It was observed already by Sutherland \cite{Suth2} that a key
property in their construction is that this dif\/ferential
operator can be consistently restricted to certain f\/inite
dimensional subspaces of the symmetric polynomials, on which it
can be represented by a f\/inite dimensional triangular matrix.
This reduces the problem of constructing the reduced
eigenfunctions of the Schr\"odinger operator \eqref{genHam} to
that of diagonalising a f\/inite dimensional triangular matrix.

To make this more precise we now present a slight modif\/ication
of Sutherland's original argument for the so-called Calogero
model, def\/ined by the Schr\"odinger operator \eqref{genHam} for
$V(x) = x^2$ and $W(x,y) = 2\kappa(\kappa - 1)(x - y)^{-2}$ with
$\kappa>0$, i.e.,
\begin{gather}
\label{calogeroHam}
  H_N = -\sum\limits_{j=1}^N\frac{\partial^2}{\partial x_j^2} +
  \sum\limits_{j=1}^Nx_j^2 + 2\kappa(\kappa - 1)\sum\limits_{j<k}\frac{1}{(x_j -
  x_k)^2}.
\end{gather}
Note that we without loss of generality have set the harmonic
oscillator frequency $\omega>0$ to $1$: it can be introduced by
scaling $x_j\rightarrow \sqrt\omega x_j$ and $H\rightarrow \omega
H$. It was shown by Calogero \cite{Calogero1,Calogero2} that this
Schr\"odinger operator has eigenfunctions of the form
\eqref{genEigFuncs}, with groundstate
\begin{gather}
\label{calGroundstate}
  \Psi_0(x_1,\ldots,x_N) =
  \prod\limits_{j=1}^N\mathrm{e}^{-({1}/{2})x_j^2}\prod\limits_{j<k}(x_k - x_j)^\kappa
\end{gather}
corresponding to the eigenvalue
\begin{gather*}
  E_0 = N(1 + \kappa(N - 1)),
\end{gather*}
and where the symmetric polynomials $P_\lambda$ are natural
many-variable generalisations of the Hermite polynomials. This
implies that these symmetric polynomials are eigenfunctions of the
dif\/ferential operator
\begin{gather}
  \tilde H_N := \Psi_0^{-1}(H_N - E_0)\Psi_0 \nonumber
  \\ \qquad
  {}=  -\sum\limits_{j=1}^N\frac{\partial^2}{\partial x_j^2} + 2
  \sum\limits_{j=1}^Nx_j\frac{\partial}{\partial x_j} -
  2\kappa\sum\limits_{j<k}\frac{1}{x_j - x_k}\biggl(\frac{\partial}{\partial
  x_j} - \frac{\partial}{\partial x_k}\biggr).
\label{reducedCalHam}
\end{gather}

The idea is now to compute the action of this dif\/ferential
operator on the so-called monomial symmetric polynomials
$m_\lambda$, for each partition $\lambda =
(\lambda_1,\ldots,\lambda_N)$ def\/ined by
\begin{gather*}
  m_\lambda(x_1,\ldots,x_N) = \sum\limits_P x^{\lambda_{P(1)}}_1\cdots
  x^{\lambda_{P(N)}}_N,
\end{gather*}
where the sum extends over all distinct permutations $P$ of the
parts $\lambda_j$ of the partition $\lambda$. In the discussion
below we will on occasion refer to monomials $m_n$ parametrised by
integer vectors $n\in\mathbb{N}_0^N$ which are not partitions.
Such a monomial is then def\/ined by the equality $m_n =
m_{p(n)}$, where $p(n)$ refers to the unique partition obtained by
permuting the parts $n_j$ of $n$. Note that as $\lambda$ runs
through all partitions of length at most $N$ the monomials
$m_\lambda$ form a linear basis for the space of symmetric
polynomials in $N$ variables. Using the fact that
\begin{gather*}
  \biggl( -\frac{\partial}{\partial x} + 2x\frac{\partial}{\partial x}\biggr)
  x^n = 2nx^n - n(n - 1)x^{n-2},
\end{gather*}
as well as the identity
\begin{gather*}
  \frac{1}{x \!-\! y}\biggl(\frac{\partial}{\partial x} \!-\!
  \frac{\partial}{\partial y}\biggr)(x^ny^m \!+\! y^nx^m) =
  (n\!-\!m)\!\!\sum\limits_{k=1}^{n\!-\!m\!-\!1}x^{n\!-\!1\!-\!k}y^{m\!-\!1\!+\!k} \!-\! m(x^{n-1}y^{m-1}
  \!+\!  y^{n-1}x^{m-1}),
\end{gather*}
valid for all $x,y\in\mathbb{R}$ and $n,m\in \mathbb{N}_0$ such
that $n\geq m$, it is straightforward to verify that
\begin{gather}
  \tilde H_N m_\lambda = 2|\lambda|m_\lambda -\! \sum\limits_{j=1}^N
  \lambda_j(\lambda_j - 1)m_{\lambda-2e_j} -\!
  2\kappa\sum\limits_{j<k}\sum\limits_{\nu=1}^{\lfloor(\lambda_j-\lambda_k)/2\rfloor}\!\!\!\!\!
  (\lambda_j - \lambda_k)m_{\lambda - (\nu+1)e_j + (\nu-1)e_k}\nonumber
  \\ \phantom{\tilde H_N m_\lambda =}{}
  + 2\kappa\sum\limits_{j<k}\lambda_k m_{\lambda-e_j-e_k},
\label{reducedAction}
\end{gather}
where $|\lambda| = \lambda +\cdots + \lambda_N$, $\lfloor
n/2\rfloor$ denotes the integer part of $n/2$, and $e_j$ are the
natural basis elements in $\mathbb{Z}^N$ def\/ined by $(e_j)_k =
\delta_{jk}$. It is important to note at this point that the right
hand side of this expression involves terms which in general are
not parametrised by partitions, e.g., if $\lambda = (3,2,2)$ then
$\lambda - 2e_1 = (1,2,2)$ which is not a partition. However,
since $m_{P(\lambda)} = m_\lambda$, for all permutations $P$ of
$N$ objects, we can remedy this problem by collecting all terms
corresponding to the same monomial $m_\lambda$. Once this is done
we f\/ind that the action of the dif\/ferential
opera\-tor~\eqref{reducedCalHam} on the monomials $m_\lambda$ is
triangular, in the sense that if two partitions $\mu =
(\mu_1,\ldots,\mu_N)$ and $\lambda = (\lambda_1,\ldots,\lambda_N)$
are ordered according to the partial ordering
\begin{gather*}
  \mu\leq\lambda~\Leftrightarrow~\mu_1 + \cdots + \mu_j\leq \lambda_1
  + \cdots + \lambda_j,\qquad \forall \, j = 1,\ldots,N,
\end{gather*}
then $\tilde H m_\lambda$ is a linear combination of $m_\lambda$
and
 monomials $m_\mu$ with $\mu<\lambda$ and $|\mu|\leq |\lambda| - 2$,
 i.e.,
\begin{gather}
\label{orderedAction}
  \tilde H_N m_\lambda = 2|\lambda|m_\lambda +
  \sum\limits_{\mu}c_{\lambda\mu}m_\mu
\end{gather}
for some coef\/f\/icients $c_{\lambda\mu}$, and where the sum is
over partitions $\mu<\lambda$ such that $|\mu|\leq |\lambda| - 2$.
This means that when constructing the reduced eigenfunctions of
the Calogero model we can restrict the dif\/ferential operator
\eqref{reducedCalHam} to a subspace of the symmetric polynomials
spanned by monomials~$m_\mu$, where $\mu\leq\lambda$ for some
f\/ixed partition $\lambda$. On this subspace the dif\/ferential
operator \eqref{reducedCalHam} can indeed be represented by a
f\/inite dimensional triangular matrix, with of\/f-diagonal
elements $c_{\lambda\mu}$, and where its diagonal elements
$2|\lambda|$ give the eigenvalues for the reduced eigenfunctions
of the Calogero model, which correspond to the eigenvectors of
this matrix. There remains then to actually compute the matrix
elements $c_{\lambda\mu}$, i.e., to collect all terms in
\eqref{reducedAction} corresponding to the same monomial $m_\mu$.
It seems however that this problem does not have a simple
solution, which in turn implies that the reduced eigenfunctions of
the Calogero model do not have a simple series representation in
terms of monomial symmetric polynomials. The situation is similar
for the other models discussed above (see e.g. \cite{HallLang2}),
and as far as we know also for other simple bases of the space of
symmetric polynomials, such as elementary, complete homogeneous
and power sum symmetric polynomials; see e.g. \cite{MacD} for
their def\/inition.

\subsection{A sketch of our solution method}
To obtain our explicit analytic series representation for the
reduced eigenfunctions of Calogero--Sutherland type models with
polynomial eigenfunctions we use a construction which dif\/fers
from the one discussed above in two important aspects: f\/irst, we
express them in terms of a particular set of symmetric polynomials
$f_n$, $n\in\mathbb{Z}^N$, on which the action of the
dif\/ferential operator \eqref{reducedGenHam} is simpler than on
the symmetric monomials; second, we avoid the problem of computing
the matrix elements analogous to the $c_{\lambda\mu}$ in
\eqref{orderedAction} by using an overcomplete set of these
polynomials, parametrised not only by partitions but by a larger
set of integer vectors in $\mathbb{Z}^N$. One could of course
apply this latter change to the symmetric monomials and the
discussion in the previous section. Note, however, that the
expression \eqref{reducedAction} for the action of the
dif\/ferential operator \eqref{reducedCalHam} on the symmetric
monomials is valid only for partitions, and that a formula valid
for arbitrary integer vectors in $\mathbb{N}_0^N$ would be more
involved.

To simplify notation we will here, and in the remainder of the
paper, let $x = (x_1,\ldots,x_N)$ and $y = (y_1,\ldots,y_N)$ be
two sets of independent variables. For an arbitrary integer vector
$n\in\mathbb{Z}^N$ we will furthermore use the notation $x^n =
x_1^{n_1}\cdots x_N^{n_N}$, and similarly for $y$. We now def\/ine
the set of symmetric polynomials $f_n$, $n\in\mathbb{Z}^N$,
through the expansion of their generating function
\begin{gather}
\label{fnDef}
  \frac{\prod\limits_{j<k}\biggl(1 - \dfrac{y_j}{y_k}\biggr)^\kappa}{\prod\limits_{j,k}
  \biggl(1 -  \dfrac{x_j}{y_k}\biggr)^\kappa} = \sum\limits_{n\in\mathbb{Z}^N}f_n(x)y^{-n},
\end{gather}
valid for $|y_N|>\cdots>|y_1|>\max_j(|x_j|)$. Although the
expansion unavoidably generates terms parametrised by integer
vectors which are not partitions, we prove in Section
\ref{completenessSection} that a basis for the space of symmetric
polynomials is formed by those $f_n$ which are parametrised by
partitions alone. The reason that we use precisely these symmetric
polynomials is that for each Schr\"odinger operator \eqref{genHam}
there exists an identity
\begin{gather}
\label{genId}
  H_N(x)F(x,y) = \big( H^{(-)}_N(y) + C_N\big) F(x,y),
\end{gather}
where $C_N$ is a constant, $H^{(-)}_N$ is obtained from $H_N$ by a
simple shift in its parameters (see \cite{HallLang2}), and the
function $F$ is given by
\begin{gather*}
  F(x,y) =
  \Psi_0(x)\prod\limits_{j=1}^N\psi^{(-)}_0(y_j)\frac{\prod\limits_{j<k}(z(y_k) -
  z(y_j))^\kappa}{\prod\limits_{j,k}(z(y_k) - z(x_j))^\kappa},
\end{gather*}
where $\psi^{(-)}_0$ is the groundstate of the one-body model
obtained by setting $N = 1$ in $H^{(-)}_N$. Note that if the
groundstate factors are removed and the variables $z_j = z(x_j)$
and $w_j = z(y_k)$ are introduced we essentially recover the
generating function for the symmetric polynomials $f_n$. This
relation will later enable us to obtain the action of the
dif\/ferential operator \eqref{reducedGenHam} on the~$f_n$ in a
straightforward manner. As is then shown, this action is simple
enough to be inverted explicitly, thus yielding our explicit
analytic series representation for the reduced eigenfunctions of
Calogero--Sutherland type models with polynomial eigenfunctions.

In the literature there exist various other approaches to the
construction of these reduced eigenfunctions. In a recent paper
Lassalle and Schlosser \cite{LassSchloss} obtained two explicit
analytic series expansions for the Jack polynomials, the reduced
eigenfunctions of the Sutherland model, by inverting their so
called Pieri formula. For very particular partitions or a low
number of variables explicit analytic expansions have also been
obtained by other methods; see e.g.~\cite{MacD}. In addition,
various representations of a combinatorial nature are known for
the Jack, as well as certain other related many-variable
polynomials \cite{DLM,KS,MacD,vDLM}. We also mention the recent
separation-of-variables approach to the Sutherland model due to
Kuznetsov, Mangazeev and Sklyanin \cite{KMS}, which also relies on
the identity \eqref{genId}. This list of previous results
ref\/lects only those which we have found to be most closely
related to ours. For a more comprehensive discussion we refer to
\cite{HallLang2}.

\subsection{An outline for the remainder of the paper}
We continue in Section 2 to give a more detailed account of our
solution method by applying it to the particular case of the
Calogero model. In Section 3 we then discuss generalisations of
this result to other Calogero--Sutherland type models with
polynomial eigenfunctions and also to the `deformed'
Calogero--Sutherland type models recently introduced and studied
by Chalykh, Feigin, Sergeev and Veselov; see
\cite{CFV,SergVes1,SergVes2} and references therein.

\section{A f\/irst example: eigenfunctions of the Calogero model}
In this section we provide a detailed account of our solution
method by applying it to the Calogero model, def\/ined by the
Schr\"odinger operator \eqref{calogeroHam}. Apart from the proof
of completeness these results were all obtained in
\cite{HallLang1}.

We begin by formulating our main result: an explicit analytic
series representation for the reduced eigenfunctions of the
Calogero model in terms of the symmetric polynomials $f_n$. In
doing so we make use of a few notational conventions which we now
introduce. In contrast to the introduction we will here use the
following partial ordering ordering of integer vectors
$m,n\in\mathbb{Z}^N$:
\begin{gather*}
  m\preceq n~\Leftrightarrow~m_j +\cdots + m_N\leq n_j +\cdots +
  n_N,\qquad \forall \, j = 1,\ldots,N.
\end{gather*}
To simplify certain formulae we associate to each
$n\in\mathbb{Z}^N$ the shifted integer vector
\begin{gather*}
  n^+ = (n_1^+,\ldots,n^+_N),\qquad n^+_j = n_j + \kappa (N + 1 - j).
\end{gather*}
For each integer vector $n\in\mathbb{Z}^N$ we def\/ine the
Kronecker delta
\begin{gather*}
  \delta_n(m) = \prod\limits_{j=1}^N \delta_{n_jm_j}.
\end{gather*}
We also recall the notation $e_j$ for the standard basis in
$\mathbb{Z}^N$, i.e., $(e_j)_k = \delta_{jk}$. We are now ready to
state the main result of this section.

\begin{theorem}\label{mainCalogeroThm}
For an arbitrary integer vector $n\in\mathbb{Z}^N$ let
\begin{gather}
\label{redCalEigfuncs}
  P_n = f_n + \sum\limits_m u_n(m)f_m,
\end{gather}
where the sum is over integer vectors $m\in\mathbb{Z}^N$ such that
\begin{gather*}
  m\prec n~\mathrm{and}~|m|\leq|n| - 2,
\end{gather*}
and the coefficients
\begin{gather*}
  u_n(m) = \sum\limits_{s=1}^\infty\frac{1}{4^s s!}\sum\limits_{j_1\leq k_1}\cdots
  \sum\limits_{j_s\leq k_s}\sum\limits_{\nu_1,\ldots,\nu_s=0}^\infty\delta_n\Biggl( m +
  \sum\limits_{r=1}^sE^{\nu_r}_{j_rk_r}\Biggr)\prod\limits_{r=1}^s
  g_{j_rk_r}\Biggl(\nu_r;n - \sum\limits_{\ell=1}^rE^{\nu_\ell}_{j_\ell
  k_\ell}\Biggr),
\end{gather*}
where we use the shorthand notation
\begin{gather}
\label{gCoeffs}
  g_{jk}(\nu;m) = 2\kappa(\kappa - 1)\nu(1 - \delta_{jk}) -
  m_j^+(m_j^+ + 1)\delta_{\nu 0}\delta_{jk}
\end{gather}
and
\begin{gather*}
  E^\nu_{jk} = (1 - \nu)e_j + (1 + \nu)e_k.
\end{gather*}
Then $P_n$ is a reduced eigenfunction of the Schr\"odinger
operator \eqref{calogeroHam} corresponding to the eigenvalue
\begin{gather*}
  E_n = 2|n| + E_0,\qquad E_0 = N(1 + \kappa(N - 1)).
\end{gather*}
Moreover, as $\lambda$ runs through all partitions of length at
most $N$ the $P_\lambda$ form a basis for the space of symmetric
polynomials in $N$ variables.
\end{theorem}

\begin{remark}
It is important to note that the series def\/ining the
coef\/f\/icients $u_n(m)$ terminate after a f\/inite number of
terms, and thus are well-def\/ined. This is a direct consequence
of the def\/inition of the Kronecker-delta $\delta_n(m)$ and the
fact that the equations
\begin{gather*}
  n - m = \sum\limits_{r=1}^sE^{\nu_r}_{j_rk_r}
\end{gather*}
only have a f\/inite number of solutions $\nu =
(\nu_1,\ldots,\nu_s)$ for f\/ixed $n,m\in\mathbb{Z}^N$.
\end{remark}

To prove the theorem we proceed in three steps: we begin by
deriving the identity \eqref{genId} for the Schr\"odinger operator
\eqref{calogeroHam}; we then prove the f\/irst part of the
theorem, that the functions $P_n$ are reduced eigenfunctions of
the Schr\"odinger operator \eqref{calogeroHam}; and f\/inally, we
prove that a basis for its eigenspace is given by those
eigenfunctions which are parametrised by a partition of length at
most $N$.

\subsection{The identity and a model with dif\/ferent masses}
Rather than proving the identity \eqref{genId} for the
Schr\"odinger operator \eqref{calogeroHam} by a direct computation
we obtain it here as a consequence of a more general result which
has the interpretation of providing the exact groundstate of a
generalisation of the Calogero model where the particles are
allowed to have dif\/ferent masses. We will, however, not stress
this interpretation but rather use the result to derive various
other identities, of which \eqref{genId} is the one of main
interest for the discussion which follows.

\begin{proposition}\label{idProp}
For a given set of real non-zero parameters $m =
(m_1,\ldots,m_\mathcal{N})$ and variables $X =
(X_1,\ldots,X_\mathcal{N})$ let
\begin{gather}\label{massHam}
  \mathcal{H} =
  -\sum\limits_{j=1}^\mathcal{N}\frac{1}{m_j}\frac{\partial^2}{\partial
  X_j^2} + \sum\limits_{j=1}^\mathcal{N} m_jX_j^2 +
  \kappa\sum\limits_{j<k}(\kappa m_jm_k - 1)(m_j + m_k)\frac{1}{(X_j-X_k)^2}
\end{gather}
and let
\begin{gather*}
  \Phi_0(X_1,\ldots,X_\mathcal{N}) =
  \prod\limits_{j=1}^\mathcal{N}\psi_{0,m_j}(X_j)\prod\limits_{j<k}(X_k - X_j)^{\kappa
  m_jm_k},\qquad \psi_{0,m_j}(X_j) = \mathrm{e}^{-m_jX_j^2/2}.
\end{gather*}
We then have that
\begin{gather}\label{groundstateId}
  \mathcal{H}\Phi_0 = \mathcal{E}_0\Phi_0
\end{gather}
with the constant
\begin{gather*}
  \mathcal{E}_0 = \kappa\Biggl(\sum^\mathcal{N}_{j=1} m_j\Biggr)^2 +
  \sum\limits_{j=1}^\mathcal{N} (1 - \kappa m_j^2).
\end{gather*}
Moreover, if all $m_j$ are positive and $\Phi_0$ is square
integrable then $\mathcal{H}$ defines a self-adjoint operator
bounded from below by $\mathcal{E}_0$ and with groundstate
$\Phi_0$.
\end{proposition}

\begin{proof}
We prove the statement by establishing that the dif\/ferential
operator \eqref{massHam} is factorisable according to
\begin{gather*}
  \mathcal{H} = \sum\limits_{j=1}^\mathcal{N}\frac{1}{m_j}Q_j^+Q_j^- +
  \mathcal{E}_0
\end{gather*}
with
\begin{gather*}
  Q_j^\pm = \pm\partial_{X_j} + \mathcal{V}_j,\qquad \mathcal{V}_j =
  \Phi_0^{-1}\partial_{X_j}\Phi_0.
\end{gather*}
Note that $Q_j^+$ is the formal adjoint of $Q_j^-$. The identity
\eqref{groundstateId} then follows from the fact that $Q_j^-\Phi_0
= 0$ for all $j$. If all $m_j$ are positive then this
factorisation shows that $\mathcal{H}$ def\/ines a unique
self-adjoint operator via the Friedrichs extension which is
bounded from below by $\mathcal{E}_0$ (see e.g. Theorem X.23 in
\cite{ReedSimon}) and with $\Phi_0$ as ground state.

Observing that
\begin{gather*}
  \mathcal{V}_j(X_1,\ldots,X_\mathcal{N}) = -m_jX_j +
  \kappa\sum\limits_{k\neq j}m_jm_k\frac{1}{X_j-X_k}
\end{gather*}
it is straightforward to deduce that
\begin{gather*}
  \sum\limits_{j=1}^\mathcal{N}\frac{1}{m_j}Q_j^+Q_j^- = \mathcal{H} -
  \mathcal{R}
\end{gather*}
with remainder term
\begin{gather*}
   \mathcal{R} = 2\kappa\sum\limits_{k\neq j}m_jm_k\frac{X_j}{X_j - X_k} +
   \kappa^2\sum\limits_{\substack{k,l\neq j\\l\neq
   k}}\frac{m_jm_km_l}{(X_k-X_j)(X_j-X_l)} + \mathcal{N}.
\end{gather*}
Upon symmetrising the double sum and using the identity
\begin{gather*}
  \frac{1}{(X_k - X_j)(X_j - X_l)} + \frac{1}{(X_l - X_k)(X_k - X_j)}
  + \frac{1}{(X_j - X_l)(X_l - X_k)} = 0
\end{gather*}
it is readily verif\/ied that
\begin{gather*}
  \mathcal{R} = \kappa\sum\limits_{k\neq j}m_jm_k + \mathcal{N} =
  \kappa\Biggl(\sum\limits_{j=1}^\mathcal{N} m_j\Biggr)^2 +
  \sum\limits_{j=1}^\mathcal{N} (1 - \kappa m_j^2) = \mathcal{E}_0.\tag*{\qed}
\end{gather*}\renewcommand{\qed}{}
\end{proof}

We note that by setting all $m_j=1$ we obtain as a direct
consequence of the proposition that \eqref{calGroundstate} indeed
is the groundstate of the Calogero model. On the other hand,
setting $\mathcal{N} = 2N$, $m_j = 1$ and $m_{N+j} = -1$ for $j =
1,\ldots,N$ we see that $\mathcal{H}$ splits into a dif\/ference
of two Schr\"odinger operators \eqref{calogeroHam} and that we
obtain the corresponding identity \eqref{genId} with $H^{(-)}_N =
H_N$.

\begin{corollary}\label{cor1}
With
\begin{gather*}
  F(x,y) = \Psi_0(x)\prod\limits_{j=1}^N\psi_{0,-1}(y_j)\dfrac{\prod\limits_{j<k}(y_k
  - y_j)^\kappa}{\prod\limits_{j,k}(y_k - x_j)^\kappa}
\end{gather*}
we have that
\begin{gather}\label{Id}
  H_N(x)F(x,y) = \left(H_N(y) + C_N\right)F(x,y),
\end{gather}
where the constant
\begin{gather*}
  C_N = 2(1 - \kappa)N.
\end{gather*}
\end{corollary}

It is interesting to observe that Proposition \ref{idProp} implies
a number of additional identities. We can for example choose to
take dif\/ferent number of variables $x_j$ and $y_k$. This leads
to an identity involving two Schr\"odinger operators $H_N$ and
$H_M$ with dif\/ferent number of variables $N$ and $M$. We may
also set some of the parameters $m_j$ to either $1/\kappa$ or
$-1/\kappa$ while still preserving the property that $\mathcal{H}$
splits into a dif\/ference of two dif\/ferential operators, which
in this case will def\/ine so-called `deformed'
Calogero--Sutherland type models; see Section \ref{defSection}.
These additional identities are further discussed in
\cite{HallLang2}.

\subsection{Construction of reduced eigenfunctions}
We proceed to prove the f\/irst part of the statement in Theorem
\ref{mainCalogeroThm}, that the symmetric polynomials $P_n$,
def\/ined by \eqref{redCalEigfuncs}, are reduced eigenfunctions of
the Schr\"odinger operator \eqref{calogeroHam}. We begin by
computing the action of the dif\/ferential operator
\eqref{reducedCalHam} on the symmetric polynomials $f_n$.

\begin{lemma}\label{actionLemma}
For each $n\in\mathbb{Z}^N$ we have that
\begin{gather}\label{action}
  \tilde H_N f_n = \tilde E_n f_n - \sum\limits_{j=1}^N(n_j^+ - 1)(n_j^+ -
  2)f_{n-2e_j} + 2\kappa(\kappa - 1)\sum\limits_{j<k}\sum\limits_{\nu=1}^\infty \nu
  f_{n - (1 - \nu)e_j - (1 + \nu)e_k}
\end{gather}
with
\begin{gather}\label{tildeEn}
  \tilde E_n = E_n - E_0 = 2|n|.
\end{gather}
\end{lemma}

\begin{proof}
We f\/irst note that the function $F$ in Corollary \ref{cor1} and
the generating function for the symmetric polynomials $f_n$ are
related as follows:
\begin{gather*}
  F(x,y) =
  \Psi_0(x)\Biggl(\prod\limits_{j=1}^N\psi_{0,-1}(y_j)y_j^{-\kappa(N+1-j)}\Biggr)
  \dfrac{\prod\limits_{j<k}\biggl(1 - \dfrac{y_j}{y_k}\biggr)^\kappa}{\prod\limits_{j,k}
  \biggl(1 - \dfrac{x_j}{y_k}\biggr)^\kappa}
\end{gather*}
with $\Psi_0$ the groundstate \eqref{calGroundstate} of the
Schr\"odinger operator \eqref{calogeroHam}. The identity
\eqref{Id} in Corollary~\ref{cor1} together with def\/initions
\eqref{reducedCalHam} and \eqref{fnDef}, of respectively the
dif\/ferential operator $\tilde H_N$ and the symmetric polynomials
$f_n$, therefore imply that
\begin{gather}\label{actionId}
  \sum\limits_{n\in\mathbb{Z}^N}\left( \tilde H_N f_n(x)\right) y^{-n^+} =
  \sum\limits_{n\in\mathbb{Z}^N}f_n(x)\left( \bar H_N + C_N - E_0\right)
  y^{-n^+},
\end{gather}
where
\begin{gather}
\bar H_N  =
  \frac{1}{\prod\limits_{j=1}^N\psi_{0,-1}(y_j)}H_N\prod\limits_{j=1}^N\psi_{0,-1}(y_j)\nonumber
  \\ \phantom{\bar H_N}
   {}= -\sum\limits_{j=1}^N\frac{\partial^2}{\partial y_j^2} -
  \sum\limits_{j=1}^N\biggl(2y_j\frac{\partial}{\partial y_j} + 1\biggr) +
  2\kappa(\kappa - 1)\sum\limits_{j<k}\frac{1}{(y_j - y_k)^2}.
\label{semiRedCalHam}
\end{gather}
We now expand the interaction term in a geometric series
\begin{gather*}
  \frac{1}{(y_j - y_k)^2} = \sum\limits_{\nu=1}^\infty\nu
  \frac{y_j^{\nu-1}}{y_k^{\nu+1}},
\end{gather*}
which in the region $|y_N|>|y_{N-1}|>\cdots >|y_1|$ is valid for
all $j<k$. It is now straightforward to compute the right hand
side of \eqref{actionId}, and by comparing coef\/f\/icients of
$y^{-n^+}$ on both sides of the resulting equation we obtain
\eqref{action} with
\begin{gather*}
  \tilde E_n = \sum\limits_{j=1}^N(2n_j^+ - 1) + C_N - E_0.
\end{gather*}
As a simple computation shows, this indeed coincides with
\eqref{tildeEn}, and the statement is thereby proved.
\end{proof}

\begin{remark}
At this point it is interesting to compare the action of the
dif\/ferential opera\-tor~\eqref{reducedCalHam} on the monomials
$m_\lambda$, given by \eqref{reducedAction}, and on the
polynomials $f_n$, as just obtained. We note, in particular, that
the simpler structure of the latter arise from the fact that it is
essentially equivalent to the action of the dif\/ferential
operator $\bar H_N$, def\/ined by \eqref{semiRedCalHam}, on the
powers $y^{-n^+}$ and the fact that the `interaction' terms of
this operator does not contain any derivatives, in contrast to the
dif\/ferential operator \eqref{reducedCalHam}.
\end{remark}

It is clear from \eqref{action} that the action of the
dif\/ferential operator \eqref{reducedCalHam} on the symmetric
polynomials $f_n$ has a triangular structure, in the sense that
$\tilde H f_n$ is a linear combination of $f_n$ and symmetric
polynomials $f_m$ with $m\prec n$ and $|m|\leq |n| - 2$. This
suggests that to each $n\in\mathbb{Z}^N$ corresponds a reduced
eigenfunction $P_n$ of the form \eqref{redCalEigfuncs} with
eigenvalue $\tilde E_n$. Inserting this ansatz into \eqref{action}
and introducing $u_n(n) = 1$ we obtain
\begin{gather*}
  \tilde H_N P_n = \tilde E_n f_n + \sum\limits_m\Bigg(\tilde E_m
  u_n(m) - \sum\limits_{j=1}^N(m_j^+ + 1)m_j^+u_n(m + 2e_j)
  \\ \phantom{\tilde H_N P_n =}
  {}+  2\kappa(\kappa - 1)\sum\limits_{j<k}\sum\limits_{\nu=1}^\infty\nu
  u_n(m+(1-\nu)e_j+(1+\nu)e_k)\Bigg)f_m,
\end{gather*}
where the sum is over integer vectors $m\in\mathbb{Z}^N$ such that
$m\prec n$ and $|m|\leq |n| - 2$. We therefore conclude that the
validity of the Schr\"odinger equation $H_N\Psi_n = E_n\Psi_n$
follows from the recursion relation
\begin{gather*}
  2(|n| - |m|)u_n(m) = \sum\limits_{j\leq k}\sum\limits_{\nu = 0}^\infty
  g_{jk}(\nu;m)u_n(m+E^\nu_{jk}),
\end{gather*}
with the coef\/f\/icients $g_{jk}(\nu;m)$ def\/ined by
\eqref{gCoeffs} and where we used the fact that $\tilde E_n -
\tilde E_m = 2(|n| - |m|)$. We now proceed to solve this recursion
relation. Suppressing the argument $m$ we rewrite it in the form
\begin{gather*}
  u_n = \delta_n + Ru_n,
\end{gather*}
where the operator $R$ is def\/ined by
\begin{gather*}
  (Ru_n)(m) = \frac{1}{2(|n| - |m|)}\sum\limits_{j\leq k}\sum\limits_{\nu=0}^\infty
  g_{jk}(\nu;m)u_n(m+E^\nu_{jk}).
\end{gather*}
Observe that this expression is well-def\/ined since $|n|-|m|\neq
0$ for all applicable $m$ and the sum truncates after a f\/inite
number of terms: $u_n(m)$ is by def\/inition non-zero only if
$m\preceq n$. The solution of this latter equation is therefore
\begin{gather*}
  u_n = (1 - R)^{-1}\delta_n = \sum\limits_{s=0}^\infty R^s\delta_n,
\end{gather*}
where the expansion into a geometric series is well-def\/ined
since it only contains a f\/inite number of non-zero terms, as
will become apparent below. Using the def\/inition of the operator
$R$ as well as the def\/ining properties of the Kronecker delta
$\delta_n$ we deduce that
\begin{gather*}
  (R^s\delta_n)(m) = \sum\limits_{j_s\leq
  k_s}\sum\limits_{\nu_s=0}^\infty\frac{g_{j_sk_s}(\nu_s;m)}{2(|n|-|m|)}\sum\limits_{j_{s-1}\leq
  k_{s-1}}\sum\limits_{\nu_{s-1}=0}^\infty\frac{g_{j_{s-1}k_{s-1}}\bigl(\nu_{s-1};m
  +  E^{\nu_s}_{j_sk_s}\bigr)}{2\bigl(|n|-\bigl|m+E^{\nu_s}_{j_sk_s}\bigr|\bigr)}\times\cdots
  \\ \phantom{(R^s\delta_n)(m) =}
{} \times\sum\limits_{j_1\leq
k_1}\sum\limits_{\nu_1=0}^\infty\frac{g_{j_1k_1}\Bigl(\nu_1;m +
  \sum\limits_{\ell=2}^sE^{\nu_\ell}_{j_\ell k_\ell}\Bigr)}{2\Bigl(|n| - \Bigl|m+\sum\limits_{\ell=2}^sE^{\nu_\ell}_{j_\ell
  k_\ell}\Bigr|\Bigr)}\delta_n\Biggl( m + \sum\limits_{r=1}^sE^{\nu_r}_{j_rk_r}\Biggr)
  \\ \phantom{(R^s\delta_n)(m)}
  {}= \sum\limits_{j_1\leq k_1}\cdots  \sum\limits_{j_s\leq k_s}\sum\limits_{\nu_1,\ldots,\nu_s=0}^\infty\delta_n\Biggl( m +
  \sum\limits_{r=1}^sE^{\nu_r}_{j_rk_r}\Biggr)\prod\limits_{r=1}^s\frac{g_{j_rk_r}\Bigl(\nu_r;n
  - \sum\limits_{\ell=1}^rE^{\nu_\ell}_{j_\ell k_\ell}\Bigr)}{2\Bigl(|n| -
  \Bigl|n - \sum\limits_{\ell=1}^rE^{\nu_\ell}_{j_\ell  k_\ell}\Bigr|\Bigr)}.
\end{gather*}
By f\/inally observing that
\begin{gather*}
  2\Biggl(|n| - \Biggl|n - \sum\limits_{\ell=1}^rE^{\nu_\ell}_{j_\ell k_\ell}\Biggr|\Biggr) = 4r
\end{gather*}
we obtain our explicit analytic series representation
\eqref{redCalEigfuncs} for the reduced eigenfunctions of the
Calogero model.

\subsection{Completeness of the reduced eigenfunctions}\label{completenessSection}
There remains only to prove that the reduced eigenfunctions just
obtained provide a basis for the space of symmetric polynomials,
i.e., that they span the eigenspace of the dif\/ferential
opera\-tor~\eqref{reducedCalHam}. We obtain this last part of
Theorem \ref{mainCalogeroThm} by exploiting the relation between
the symmetric polynomials $f_n$ and the so-called `modif\/ied
complete' symmetric polynomials $g_\lambda$, def\/ined through the
expansion of their generating function
\begin{gather*}
  \frac{1}{\prod\limits_{j,k}\biggl(1 - \dfrac{x_j}{y_k}\biggr)^\kappa} = \sum\limits_\lambda
  g_\lambda(x)m_\lambda(y^{-1}),
\end{gather*}
valid for $\min_k|y_k|>\max_j(|x_j|)$, and where the summation
extends over all partitions of length at most $N$. It is well
known that the $g_\lambda$ are homogeneous symmetric polynomials
of degree $|\lambda|$, and also that as $\lambda$ runs through all
partitions of length at most $N$ they form a basis for the space
of symmetric polynomials in $N$ variables; see e.g. Section VI.10
in \cite{MacD}. We mention that these f\/irst properties can be
directly inferred from their generating function, whereas the fact
that they span the space of symmetric polynomials is a consequence
of the equivalence between the expansion by which they are
def\/ined and the fact that they are dual to the to the monomial
symmetric polynomials $m_\lambda$ in a particular inner product;
see e.g. Statement 10.4 in \cite{MacD}.

By comparing the generating functions for the $f_n$ and the
$g_\lambda$ we f\/ind that
\begin{gather*}
  \sum\limits_n f_n(x) y^{-n} = \prod\limits_{j<k}\biggl(1 - \frac{y_j}{y_k}\biggr)^\kappa\sum\limits_\lambda
  g_\lambda(x)m_\lambda(y^{-1}).
\end{gather*}
Assuming $|y_N|>\cdots>|y_1|$ and expanding each term in the
product in a power series we rewrite the right hand side as
follows:
\begin{gather*}
  \sum\limits_{n\in\mathbb{N}_0^N}g_{p(n)}(x)\prod\limits_{j<k}\sum\limits_{p_{jk}=0}^\infty
  (-1)^{p_{jk}}\binom{\kappa}{p_{jk}}y^{-n +
  \sum\limits_{j<k}p_{jk}(e_j-e_k)},
\end{gather*}
where we have taken $p(n)$ to denote the unique partition obtained
by reordering the parts $n_j$ of~$n$. This means that
\begin{gather}\label{fing}
  f_n = \prod\limits_{j<k}\sum\limits_{p_{jk}=0}^\infty
  (-1)^{p_{jk}}\binom{\kappa}{p_{jk}}g_{p(n+\sum\limits_{j<k}p_{jk}(e_j-e_k))}.
\end{gather}
Many of the properties of the `modif\/ied complete' symmetric
polynomials $g_\lambda$ for this reason carry over to the $f_n$.
In particular, observing that
\begin{gather*}
  \Biggl|n + \sum\limits_{j<k}p_{jk}(e_j - e_k)\Biggr| = |n|
\end{gather*}
for all integers $p_{jk}$ we conclude that they are homogeneous
symmetric polynomials of degree $n$. We also see that they are
non-zero only if $n\succeq 0$. Now suppose $\lambda$ is a
partition. It is then clear that
\begin{gather*}
  m := \lambda + \sum\limits_{j<k}p_{jk}(e_j - e_k)\preceq\lambda
\end{gather*}
and furthermore that also $p(m)\preceq\lambda$: we obtain the
partition $p(m)$ from $m$ by some permutation of its parts $m_j$,
and since by def\/inition $\mu_1\geq\cdots\geq\mu_N$ for any
partition $\mu = (\mu_1,\ldots,\mu_N)$ we have that $p(m)\preceq
m\preceq\lambda$. We therefore conclude that
\begin{gather}\label{fgrel}
  f_\lambda = g_\lambda + \sum\limits_\mu M_{\lambda\mu}g_\mu
\end{gather}
for some coef\/f\/icients $M_{\lambda\mu}$, and where the sum is
over partitions $\mu\prec\lambda$. As indicated in this expression
we let $M = (M_{\lambda\mu})$ denote the transition matrix,
def\/ined by the equality $f_\lambda = \sum\limits_\mu
M_{\lambda\mu}g_\mu$, from the $f_\lambda$ to the $g_\mu$. Given a
partition $\lambda$ it follows from \eqref{fgrel} that it can be
consistently restricted to the partitions $\mu$ such that
$\mu\preceq\lambda$. With rows and columns ordered in descending
order this restricted transition matrix is upper triangular with
$1$'s on the diagonal. Hence, it can be inverted. Since the
inverse of an upper triangular matrix is upper triangular we
obtain that
\begin{gather*}
  g_\lambda = f_\lambda + \sum\limits_\mu (M^{-1})_{\lambda\mu}f_\mu,
\end{gather*}
where the sum is over partitions $\mu\prec\lambda$. We have
thereby proved the following:

\begin{proposition}\label{fProp}
The functions $f_n$ are non-zero only if $n\succeq 0$. In that
case, $f_n$ is a homogeneous symmetric polynomial of degree $|n|$.
Moreover, as $\lambda$ runs through all partitions of length at
most~$N$ the $f_\lambda$ form a basis for the space of symmetric
polynomials in $N$ variables.
\end{proposition}

The same line of reasoning can now be applied to the reduced
eigenfunctions
\begin{gather*}
  P_\lambda = f_\lambda + \sum\limits_m u_\lambda(m) f_m
\end{gather*}
which are parametrised by partitions $\lambda$. Recall that the
sum is over integer vectors $m\prec\lambda$. Using formula
\eqref{fing} and following the subsequent discussion we f\/ind
that
\begin{gather*}
  P_\lambda = g_\lambda + \sum\limits_\mu b_{\lambda\mu}g_\mu
\end{gather*}
for some coef\/f\/icients $b_{\lambda\mu}$, and where the sum now
extends only over partitions $\mu\prec\lambda$. Applying the
arguments leading up to Proposition \ref{fProp} we conclude that
this expression can be inverted to yield each $g_\lambda$ as a
linear combination of the $P_\mu$ with $\mu\preceq\lambda$. Hence,
as $\lambda$ runs through all partitions of length at most $N$ the
$P_\lambda$ form a basis for the space of symmetric polynomials in
$N$ variables. This concludes the proof of Theorem
\ref{mainCalogeroThm}.

\section{Generalisations to other models}
In this section we indicate how the results on the Calogero model
obtained in the previous section can be generalised to similar
models with polynomial eigenfunctions, including not only models
of Calogero--Sutherland type but also the `deformed'
Calogero--Sutherland models introduced and studied by Chalykh
et.al.; see \cite{CFV,SergVes1,SergVes2} and references therein. A
detailed account of these results is in preparation
\cite{HallLang2}.

\subsection[Calogero-Sutherland models with polynomial eigenfunctions]{Calogero--Sutherland models with polynomial eigenfunctions}

When the number of particles are set to one in the Calogero model
it reduces to the very well known harmonic oscillator, which
features eigenfunctions given by the classical Hermite
polynomials. This is only one special case of the following well
known and more general statement: to each complete sequence of polynomials
$\lbrace p_n:n\in\mathbb{N}_0\rbrace$, obeying a second order
ordinary dif\/ferential equation, there is a corresponding
Schr\"odinger operator
\begin{gather}\label{genOneHam}
  h = -\frac{\partial^2}{\partial x^2} + V(x)
\end{gather}
with a particular potential function $V$ such that its
eigenfunctions are of the form
\begin{gather*}
  \psi_n(x) = \psi_0(x)p_n(z(x))
\end{gather*}
for some functions $\psi_0$ and $z$. This can be verif\/ied by
f\/irst observing that such a set of polynomials are
eigenfunctions of a dif\/ferential operator
\begin{gather*}
  \tilde h = \alpha(z)\frac{\partial^2}{\partial z^2} +
  \beta(z)\frac{\partial}{\partial z},
\end{gather*}
where
\begin{gather*}
  \alpha(z) = \alpha_2 z^2 + \alpha_1 z + \alpha_0\qquad
  \mathrm{and}\qquad \beta(z) = \beta_1 z + \beta_0
\end{gather*}
for some coef\/f\/icients $\alpha_j$ and $\beta_j$. Now
introducing the variable $x = x(z)$ as a solution of the
dif\/ferential equation
\begin{gather*}
  x^\prime(z) = \frac{1}{\sqrt{\alpha(z)}}
\end{gather*}
and def\/ining the function $\psi_0$ by
\begin{gather*}
  \psi_0(x) = \mathrm{e}^{-w(z(x))},\qquad w^\prime =
  \frac{\alpha^\prime - 2\beta}{4\alpha},
\end{gather*}
it is straightforward to verify that a Schr\"odinger operator $h$
of the form \eqref{genOneHam} is obtained by conjugation of
$\tilde h$ by the function $\psi_0$ and changing the independent
variable to $x$, or to be more precise,
\begin{gather*}
  h = -\psi_0\tilde h\psi_0^{-1} = -\frac{\partial^2}{\partial x^2} +
  V(x),
\end{gather*}
with potential function
\begin{gather*}
  V(x) = v(z(x)),\qquad v = \frac{(2\beta - \alpha^\prime)(2\beta -
  3\alpha^\prime)}{16\alpha} + \frac{1}{4}\alpha^{\prime\prime} -
  \frac{1}{2}\beta^\prime.
\end{gather*}
To illustrate this general discussion we have listed the
particular values of the coef\/f\/icients $\alpha_j$ and
$\beta_j$, as well as associated functions $\psi_0$ and $z$, which
correspond to the classical orthogonal polynomials (of Hermite,
Laguerre and Jacobi) and the generalised Bessel polynomials in
Table~\ref{table}. In fact, we can by simple translations and
rescalings always reduce to one of these four cases.

\begin{table}[hbtp]
    \begin{center}
        \begin{tabular}{|c|c|c|c|c|}
            \hline
            & & & &\\[-1.5ex]
            $p_n(z)$ & $\alpha(z)$ & $\beta(z)$ & $\psi_0(x)$ & $z(x)$
            \\[1.2ex]
            \hline
            & & & &\\[-1.8ex]
            $H_n(z)$ & $1$ & $-2z$ & $\mathrm{e}^{-{x^2}/{2}}$ & $x$\\
            (Hermite) & & & &\\[0.5ex]
            $L^{(a)}_n$ & $z$ & $a + 1 - z$ & $x^a\mathrm{e}^{-{x^2}/{2}}$ & $x^2$\\
            (Laguerre) & & & &\\[0.5ex]
            $P^{(a,b)}_n(z)$ & $1 - z^2$ & $b - a - (a + b + 2)z$
            & $\sin^{a+{1}/{2}}\biggl(\dfrac{x}{2}\biggr)\cos^{b+{1}/{2}}\biggl(\dfrac{x}{2}\biggr)$ & $\cos x$\\[-1ex]
            (Jacobi) & & & &\\[0.5ex]
            $y_n(z;1-2a,2b)$ & $z^2$ & $2b + (1 - 2a)z$ & $\text{exp}(-b\text{e}^{-x} - ax)$ & $\text{e}^x$\\
            (gen. Bessel) & & & &\\[0.5ex]
            \hline
        \end{tabular}

        \caption{The particular values of coef\/f\/icients
        $\alpha_j$ and $\beta_j$, as well as associated
        functions $\psi_0$ and $z$, corresponding to the
        classical orthogonal polynomials (of Hermite, Laguerre
        and Jacobi) and the generalised Bessel polynomials.}
        \label{table}
    \end{center}\vspace{-3mm}
\end{table}

We mention that transformations of dif\/ferential equations of the
type described above are frequently used in the theory of ordinary
dif\/ferential equations of second order; see e.g.\ Section~1.8 in
Szeg\"o's classical book \cite{Sze} on orthogonal polynomials. It
is interesting to note that this transformation has a simple and
direct generalisation to many variables. Setting the interaction
potential
\begin{gather*}
  W(x,y) = \frac{\alpha(z(x)) + \alpha(z(y))}{(z(x) - z(y))^2}
\end{gather*}
and keeping the potential function $V$ as given above one
verif\/ies that the Schr\"odinger opera\-tor~\eqref{genHam} after
a conjugation by the function $\Psi_0$, as def\/ined in
\eqref{genGroundstate}, and a change of independent variables from
the $x_i$ to the $z_i$, as def\/ined above, is transformed into
the dif\/ferential operator
\begin{gather*}
  \tilde H_N  = -\Psi_0^{-1}(H - E_0)\Psi_0
  \\ \phantom{\tilde H_N}
  {}= \sum\limits_{j=1}^N\alpha(z_j)\frac{\partial^2}{\partial z_j^2} +
  \sum\limits_{j=1}^N\beta(z_j)\frac{\partial}{\partial z_j} +
  2\kappa\sum\limits_{j<k}\frac{1}{z_j -
  z_k}\biggl(\alpha(z_j)\frac{\partial}{\partial z_j} -
  \alpha(z_k)\frac{\partial}{\partial z_k}\biggr).
\end{gather*}
Following the discussion in Section 1.1 it is straightforward to
verify that the action of this dif\/ferential operator on the
monomial symmetric polynomials is triangular in the very same
ordering as in the case of the Calogero model. This means that the
Schr\"odinger operator \eqref{genHam} for these choices of
potential functions $V$ and $W$, up to degeneracies in its
spectrum, has a complete set of reduced eigenfunctions given by
symmetric polynomials. We mention that this unifying point of view
on Calogero--Sutherland type models with polynomial eigenfunctions
seems to have been little used in the literature, with the notable
exception of Gomez-Ullate, Gonz\'alez-L\'opez and Rodriguez
\cite{GUGL} who, among other things, used this point of view to
obtain the spectrum of all these models.

In \cite{HallLang2} we show that our construction of an explicit
series representation for the reduced eigenfunctions, presented in
the previous section for the Calogero model, goes through
virtually unchanged for all these models. In particular, we
generalise Theorem \ref{mainCalogeroThm} to the following:

\begin{theorem}\label{genThm}
For $n\in\mathbb{Z}^N$, the reduced eigenfunctions of the
Schr\"odinger operator \eqref{genHam} are formally given by
\begin{gather*}
  P_n = f_n + \sum\limits_m u_n(m)f_m,
\end{gather*}
where the sum is over integer vectors $m\in\mathbb{Z}^N$ such that
\begin{gather*}
  m\prec n~\mathrm{and}~|m|\leq |n| + \mathrm{deg}(\alpha) - 2,
\end{gather*}
and the coefficients
\begin{gather*}
  u_n(m) \!=\! \sum\limits_{l=1}^\infty \sum\limits_{j_1\leq k_1}\!\cdots\!\sum\limits_{j_l\leq
  k_l}\sum\limits_{p_1,\ldots,p_l=0}^2\sum\limits_{\nu_1,\ldots,\nu_l=1}^\infty
\!\!\!\!\!\delta_n\Biggl(\! m \!+\!
  \sum\limits_{t=1}^lE_{j_tk_t}^{p_t\nu_t}\Biggr)\prod\limits_{r=1}^l
  \frac{g_{j_rk_r}\Bigl(p_r,\nu_r;n\!-\!\sum\limits_{q=r}^lE_{j_qk_q}^{p_q\nu_q}\Bigr)}
  {b_n\Bigl(n\!-\!\sum\limits_{q=r}^lE_{j_qk_q}^{p_q\nu_q}\Bigr)},
\end{gather*}
where we use the shorthand notation
\begin{gather*}
  b_n(m) = \tilde E_n - \tilde E_m,\qquad \tilde E_n =
  -\sum\limits_{j=1}^N(\alpha_2 n_j(n_j - 1) + (\beta_1 + 2\kappa(N-j))n_j),
\\
  g_{jk}(p,\nu;m) = (1 - \delta_{jk})\kappa(\kappa - 1)\alpha_p(2\nu -  p)
\\  \phantom{g_{jk}(p,\nu;m) =}
{}- \delta_{jk}\delta_{\nu 1}m_j^+\big(\delta_{p0}\alpha_0(m_j^+
  + 1) + \delta_{p1}(\alpha_1(m_j^+ + \kappa + 1) - \beta_0)\big)
\end{gather*}
and
\begin{gather*}
    E_{jk}^{p\nu} = (1 - \nu)e_j + (1 - p + \nu)e_k.
\end{gather*}
If $b_n(m) \neq 0$ for all integer vectors $m\in\mathbb{Z}^N$ such
that $m\prec n$ and $|m|\leq |n| + \mathrm{deg}(\alpha) - 2$ then
$P_n$ is a well defined symmetric polynomial. Moreover, if this is
the case for all integer vectors $n\in\mathbb{Z}^N$ such that $n =
\lambda$ for some partition $\lambda$ of length at most $N$ then
the corresponding $P_\lambda$ form a linear basis for the space of
symmetric polynomials in $N$ variables.
\end{theorem}

\begin{remark}
It is important to note that the condition $b_n(m) \neq 0$,
$m\prec n$ and $|m|\leq |n| + \mathrm{deg}(\alpha) - 2$, is
essential in order for the coef\/f\/icients $u_n(m)$ to be well
def\/ined. For generic choices of the parameter $\kappa$ and the
polynomials $\alpha$ and $\beta$ it is satisf\/ied for all
$n\in\mathbb{Z}^N$; see \cite{HallLang2} for a further discussion
of this point.
\end{remark}

\begin{remark}
At this point it is interesting to enquire whether our basis for
the reduced eigenfunctions of the Schr\"odinger operator
\eqref{genHam}, as stated in Theorem \ref{genThm}, in applicable
cases stand in a simple relation to the generalised hypergeometric
polynomials of Lassalle \cite{Lass1,Lass2,Lass3} and MacDonald
\cite{MacD2}, def\/ined by expansions in Jack polynomials. In the
case $\alpha = -z^2$ and $\beta = -z$, corresponding to the Jack
polynomials themselves, one can show that they in fact coincide
and it seems natural to expect this to be true also for the
generalised Jacobi polynomials. Since the generalised Hermite and
Laguerre polynomials are limiting cases of the generalised Jacobi
polynomials (see e.g. \cite{BF}) this would imply the equivalence
also in these two cases. If established, this result would in
these cases imply a natural orthogonality for the reduced
eigenfunctions in Theorem \ref{genThm}. At this point these
statements are only conjectures and we hope to return to them
elsewhere.
\end{remark}

\subsection[Deformed Calogero-Sutherland models]{Deformed Calogero--Sutherland models}\label{defSection}

There exist an interesting deformation of Calogero--Sutherland
type models \cite{CFV,SergVes1,SergVes2}, def\/ined by the
following class of dif\/ferential operators in two sets of
variables $x = (x_1,\ldots,x_N)$ and $\tilde x = (\tilde
x_1,\ldots,\tilde x_{\tilde N})$:
\begin{gather*}
  H_{N,\tilde N} = \sum\limits_{j=1}^N\biggl( -\frac{\partial^2}{\partial
  x_j^2} + V(x_j)\biggr) - \sum\limits_{J=1}^{\tilde N}\kappa\biggl(
  -\frac{\partial^2}{\partial \tilde x_J^2} + \tilde V(\tilde x_J)\biggr)
  \\ \phantom{H_{N,\tilde N} =}
  {}+ \kappa(\kappa - 1)\sum\limits_{j<k}W(x_j,x_k) + (1 -
  \kappa)\sum\limits_{j,K}W(x_j,\tilde x_K) + \frac{\kappa -
  1}{\kappa}\sum\limits_{J<K}W(\tilde x_J,\tilde x_K),
\end{gather*}
where the potential function $\tilde V$ is obtained from $V$ by a
simple parameter shift; see \cite{HallLang2}. They provide a
natural generalisation of the models discussed in the previous
section, in that they also have polynomial eigenfunctions. To be
more precise, they have eigenfunctions which can be labelled by
partitions $\lambda = (\lambda_1,\lambda_2,\ldots)$ such that
$\lambda_{N+1}\leq \tilde N$ (see \cite{SergVes2}), and are of the
form
\begin{gather*}
  \Psi_\lambda(x,\tilde x) = \Psi_0(x,\tilde x)P_\lambda\left(
  z(x_1),\ldots,z(x_N),z(\tilde x_1),\ldots,z(\tilde x_{\tilde
  N})\right),
\end{gather*}
where the function $\Psi_0$ is given by
\begin{gather*}
  \Psi_0(x,\tilde x) = \prod\limits_{j=1}^N\psi_0(x_j)\prod\limits_{J=1}^{\tilde
  N}\tilde\psi_0(\tilde x_J)\frac{\prod\limits_{j<k}(z(x_k) -
  z(x_j))^\kappa\prod\limits_{J<K}(z(\tilde x_K) - z(\tilde
  x_J))^{1/\kappa}}{\prod\limits_{j,K}(z(\tilde x_K) - z(x_j))},
\end{gather*}
and the $P_\lambda$ are polynomials in the variables $z_j =
z(x_j)$ and $\tilde z_J = z(\tilde x_J)$. They are however no
longer symmetric under permutations of all variables but only
under permutations restricted to the $x_j$ or the $\tilde x_J$. In
addition, they obey the condition
\begin{gather*}
  \left(\frac{\partial}{\partial z_j} + \kappa\frac{\partial}{\partial
  \tilde z_J}\right) P_\lambda = 0
\end{gather*}
on the hyperplanes $z_j = \tilde z_J$, for all $j = 1,\ldots,N$
and $J = 1,\ldots,\tilde N$. The corresponding algebra of
polynomials has been extensively studied by Sergeev and Veselov
\cite{SergVes1,SergVes2}.

In \cite{HallLang2} we also construct explicit series
representations for the reduced eigenfunctions $P_\lambda$ of
these `deformed' Calogero--Sutherland type models. The
construction is analogous to the one discussed in previous
sections, with the dif\/ference that the reduced eigenfunctions
now are expressed in a set of polynomials $f_{n,\tilde{n}}$,
$(n,\tilde n)\in\mathbb{Z}^{N+\tilde N}$, def\/ined through the
expansion of their generating function
\begin{gather*}
\frac{\prod\limits_{j<k}\biggl(1 -
\dfrac{y_j}{y_k}\biggr)^\kappa\prod\limits_{J<K} \biggl(1 -
\dfrac{\tilde y_J}{\tilde
y_K}\biggr)^{1/\kappa}}{\prod\limits_{j,K} \biggl(1 -
\dfrac{y_j}{\tilde y_K}\biggr)}\frac{\prod\limits_{j,K} \biggl(1 -
\dfrac{x_j}{\tilde y_K}\biggr)\prod\limits_{J,k} \biggl(1 -
\dfrac{\tilde x_J}{y_k}\biggr)}{\prod\limits_{j,k} \biggl(1 -
\dfrac{x_j}{y_k}\biggr)^\kappa\prod\limits_{J,K} \biggl(1 -
\dfrac{\tilde x_J}{\tilde y_K}\biggr)^{1/\kappa}}
\\ \qquad
{}=  \sum\limits_{(n,\tilde{n})\in\mathbb{Z}^{N+\tilde
  N}}f_{n,\tilde{n}}(x,\tilde x)y^{-n}\tilde y^{-\tilde{n}},
\end{gather*}
valid for $|\tilde y_{\tilde N}|>\cdots|\tilde
y_1|>|y_N|>\cdots>|y_1|>\max_{j,J}(|x_j|,|\tilde x_J|)$. We
mention f\/inally that the number of variables $x_j$ and $\tilde
x_J$ as well as $y_j$ and $\tilde y_J$ may be chosen dif\/ferently
in the def\/inition of these polynomials $f_{n,\tilde n}$, thus
allowing for a number of series representations to be obtained for
the same reduced eigenfunction, an aspect of our construction
which is further discussed in~\cite{HallLang2}.

\subsection*{Acknowledgements}
I would like to thank Edwin Langmann and the three referees for a
number of helpful comments on the manuscript. Financial support
from the Knut and Alice Wallenberg foundation and the European
Union through the FP6 Marie Curie RTN {\em ENIGMA} (Contract
number MRTN-CT-2004-5652) is also gratefully acknowledged.

\newpage

\pdfbookmark[1]{References}{ref}
\LastPageEnding


\begin{thebibliography}{99}

\footnotesize\itemsep=0pt

\bibitem{BF} Baker T.H., Forrester P.J., The Calogero--Sutherland model
and generalized classical polynomials, {\it Comm. Math. Phys.}
\textbf{188} (1997), 175--216,
\href{http://arxiv.org/abs/solv-int/9608004}{solv-int/9608004}.

\bibitem{Calogero1} Calogero F., Groundstate of a one-dimensional
$N$-body system, {\it J.~Math. Phys.} \textbf{10} (1969),
2197--2200.

\bibitem{Calogero2} Calogero F., Solution of the one-dimensional $N$ body
problems with quadratic and/or inversely quadratic pair
potentials, {\it J.~Math. Phys.} \textbf{12} (1971), 419--436.

\bibitem{CFV} Chalykh O., Feigin M., Veselov A., New integrable
generalizations of Calogero--Moser quantum problems, {\it J.~Math.
Phys.} \textbf{39} (1998), 695--703.

\bibitem{DLM} Desrosiers P., Lapointe L., Mathieu P., Explicit formulas
for the generalized Hermite polynomials in superspace, {\it J.
Phys. A: Math. Gen.} \textbf{37} (2004), 1251--1268,
\href{http://arxiv.org/abs/hep-th/0309067}{hep-th/0309067}.

\bibitem{GUGL} G\'omez-Ullate D., Gonz\'alez-L\'opez A.,
Rodr{\'\i}guez M.A., New algebraic quantum many-body problems,
{\it J.~Phys.~A: Math. Gen.} \textbf{33} (2000), 7305--7335,
\href{http://arxiv.org/abs/nlin.SI/0003005}{nlin.SI/0003005}.

\bibitem{HallLang1} Halln\"as M., Langmann E., Explicit formulae for
the eigenfunctions of the $N$-body Calogero model, {\it
J.~Phys.~A: Math. Gen.} \textbf{39} (2006), 3511--3533,
\href{http://arxiv.org/abs/math-ph/0511040}{math-ph/0511040}.

\bibitem{HallLang2} Halln\"as M., Langmann E., Quantum
Calogero--Sutherland type models and generalised classical polynomials, in preparation.

\bibitem{KS} Knop F., Sahi S., A recursion and a combinatorial formula
for Jack polynomials, {\it Invent. Math.} \textbf{128} (1997),
9--22, \href{http://arxiv.org/abs/q-alg/9610016}{q-alg/9610016}.

\bibitem{KMS} Kuznetsov V.B., Mangazeev V.V., Sklyanin E.K., $Q$-operator
and factorised separation chain for Jack polynomials, {\it Indag.
Math. (N.S.)} \textbf{14} (2003), 451--482,
\href{http://arxiv.org/abs/math.CA/0306242}{math.CA/0306242}.

\bibitem{Lang1} Langmann E., Algorithms to solve the (quantum)
Sutherland model, {\it J.~Math. Phys.} \textbf{42} (2001),
4148--4157,
\href{http://arxiv.org/abs/math-ph/0104039}{math-ph/0104039}.

\bibitem{Lang2} Langmann E., A method to derive explicit formulas for
an elliptic generalization of the Jack polynomials, in Proceedings
of the Conference ``Jack, Hall-Littlewood and Macdonald
Polynomials'' (September 23--26, 2003, Edinburgh), Editors
V.B.~Kuznetsov  and S.~Sahi, {\it Contemp. Math.} \textbf{417}
(2006), 257--270,
\href{http://arxiv.org/abs/math-ph/0511015}{math-ph/0511015}.

\bibitem{Lass1} Lassalle M., Polyn\^omes de Jacobi g\'en\'eralis\'es,
{\it C. R. Math. Acad. Sci. Paris} \textbf{312} (1991), 425--428.

\bibitem{Lass2} Lassalle M., Polyn\^omes de Laguerre
g\'en\'eralis\'es, {\it C. R. Math. Acad. Sci. Paris} \textbf{312}
(1991), 725--728.

\bibitem{Lass3} Lassalle M., Polyn\^omes de Hermite g\'en\'eralis\'es,
{\it C. R. Math. Acad. Sci. Paris} \textbf{313} (1991), 579--582.

\bibitem{LassSchloss} Lassalle M., Schlosser M., Inversion of the Pieri
formula for MacDonald polynomials, {\it Adv. Math.} \textbf{202}
(2006), 289--325,
\href{http://arxiv.org/abs/math.CO/0402127}{math.CO/0402127}.

\bibitem{MacD} MacDonald I.G., Symmetric functions and Hall
polynomials, 2nd ed., Oxford University Press, 1995.

\bibitem{MacD2} MacDonald I.G., Hypergeometric functions, unpublished
manuscript.

\bibitem{OlsPer} Olshanetsky M.A., Perelomov A.M., Quantum integrable
systems related to Lie algebras, {\it Phys. Rep.} \textbf{94}
(1983), 313--404.

\bibitem{ReedSimon} Reed M., Simon B., Methods of modern mathematical
physics. II. Fourier analysis, self-adjointness, Academic Press,
1975.

\bibitem{SergVes1} Sergeev A.N., Veselov A.P., Deformed quantum
Calogero--Moser problems and Lie superalgebras, {\it Comm. Math.
Phys.} \textbf{245} (2004), 249--278,
\href{http://arxiv.org/abs/math-ph/0303025}{math-ph/0303025}.

\bibitem{SergVes2} Sergeev A.N., Veselov A.P., Generalised
discriminants, deformed Calogero--Moser--Sutherland operators and
super-Jack polynomials, {\it Adv. Math.} \textbf{192} (2005),
341--375,
\href{http://arxiv.org/abs/math-ph/0307036}{math-ph/0307036}.

\bibitem{Suth1} Sutherland B., Quantum many-body problem in one
dimension: ground state, {\it J.~Math. Phys.} \textbf{12} (1971),
246--250.

\bibitem{Suth2} Sutherland B., Exact results for a quantum many-body
problem in one dimension: II, {\it Phys. Rev. A} \textbf{5}
(1972), 1372--1376.

\bibitem{Sze} Szeg\"o G., Orthogonal polynomials, American Mathematical
Society, 1939.

\bibitem{vanDie} van Diejen J.F., Conf\/luent hypergeometric orthogonal
polynomials related to the rational quantum Calogero system with
harmonic conf\/inement, {\it Comm. Math. Phys.} \textbf{188}
(1997), 467--497,
\href{http://arxiv.org/abs/q-alg/9609032}{q-alg/9609032}.

\bibitem{vDLM} Lapointe L., Morse J., van Diejen J.F., Determinantal
construction of orthogonal polynomials associated with root
systems, {\it Compos. Math.} \textbf{140} (2004), 255--273,
\href{http://arxiv.org/abs/math.CO/0303263}{math.CO/0303263}.

\end{thebibliography}
\end{document}